\documentstyle[prl,aps,multicol]{revtex}
\input epsf

\begin{document}
\title{The Paradoxes  of the  Interaction-Free Measurements}

\author{ L. Vaidman}

\maketitle
\vspace{.4cm}
 
\centerline{  Centre for Quantum Computation}
\centerline{ Department of Physics, University of Oxford,}
\centerline{ Clarendon Laboratory, Parks Road, Oxford OX1 3PU, England.}
 \vskip .2cm
\centerline{and}
 \vskip .2cm
\centerline{ School of Physics and Astronomy}
\centerline{Raymond and Beverly Sackler Faculty of Exact Sciences}
\centerline{Tel-Aviv University, Tel-Aviv 69978, Israel}

\date{}

\vspace{.2cm}
\begin{abstract}
  Interaction-free measurements introduced by Elitzur and Vaidman
  [Found.  Phys.  {\bf 23}, 987 (1993)] allow finding infinitely
  fragile objects without destroying them. Paradoxical features of
  these and related measurements are discussed.  The resolution of the
  paradoxes in the framework of the Many-Worlds Interpretation is
  proposed.
\end{abstract}
\vspace {.1cm}

\begin{multicols}{2}

\section{INTRODUCTION}
\label{intr}

The interaction-free measurements proposed by Elitzur and Vaidman
\cite{EV91,EV93} (EV IFM) led to numerous investigations and several
experiments have been performed
\cite{Kwi95,Voo97,HaSu97,PaPa,Tse98,White98,Kwi98,seem-ifm,MiMi99,Jang,Ino,Krenn,Rud,Pott,Serge}.
Interaction-free measurements are very paradoxical. Usually it is
claimed that quantum measurements, in contrast to classical
measurements, invariably cause a disturbance of the system. The IFM is
an example of the opposite: this is a quantum measurement which does
not lead to any disturbance, while its classical counterpart
invariably does.

There are many ways to understand the interaction-free nature of the
EV IFM. A detailed analysis of various interpretations appears
elsewhere \cite{IFMifm,meaIFM}.  In this paper I will concentrate on
the paradoxical aspects of interaction-free measurements.  In Section
II, I describe the IFM of Renninger \cite{Renn} and Dicke \cite{Dick}:
changing the quantum state of a system without interaction.  In Section
III the original proposal of Elitzur and Vaidman is presented and the
basic paradox of the EV IFM is discussed: a particular interaction
leads to an explosion, nevertheless, it can be used for obtaining
information without the explosion. Section IV is devoted to another
paradoxical feature of the EV IFM: obtaining information about a region
in space without anything coming in, out, or through this place. It
also includes a brief analysis of the ``delayed choice experiment''
proposed by Wheeler \cite{Whee} which helps to define the context in
which the above claims, that the measurements are interaction-free,
are legitimate.  Section V is devoted to the variation of the EV IFM
proposed by Penrose \cite{Pen} which, instead of testing for the
presence of an object in a particular place, tests a certain property
of the object in an interaction-free way.  Section VI introduces the
EV IFM procedure for a quantum object being in a superposition of
different locations. It works equally well: it collapses the spatial
quantum state of the object to a particular place without any
disturbance of its internal state. However, the second paradoxical
feature of the EV IFM, i.e. the fact that nothing has been in the
vicinity of the object the presence of which was discovered, has a
subtle constraint. This point is explained in Section VII via the
analysis of Hardy's paradox \cite{Hardy}.  I conclude the paper in
Section VIII by arguing that the paradoxes of IFM disappear in the
framework of the many-worlds interpretation.

I want to mention a naive paradox which I have heard several times and
which I do not discuss in this paper (I discussed it elsewhere
\cite{IFMifm}).  Finding the position of a particle in an interaction
free way means, in particular, (according to these arguments) finding it without
changing its momentum. Thus, a high precision experiment of this kind
performed on a particle with bounded momentum uncertainty leads to
breaking the Heisenberg uncertainty relation. This type of arguments
appear to be due to the misleading identification of the EV IFM with an
experiment without momentum (energy) transfer
\cite{en-ex1,en-ex2,SiPl}.

\section{ The IFM of Renninger and Dicke: negative results experiment}
\label{Re-Di}

The paradox of the Renninger-Dicke type measurement is that it causes
some changes in the state of the system ``without interaction.''
Renninger discussed a {\it negative result experiment}: a situation in
which the detector does not detect anything.  In spite of the fact
that nothing happened to the detector, there is a change in the
measured system. He considered a spherical wave of a photon after it
extended beyond the radius at which a scintillation detector was located
in  part of the solid angle, see Fig.~1. The state of the detector
remained unchanged but, nevertheless, the wave-function of the photon
is modified. The name ``interaction-free'' for Renninger's setup might
be justified because there is not {\it any}, not even an infinitesimally
small, change in the state of the detector in the described process.
This is contrary to the classical physics in which interaction in
a measurement process can be made arbitrary small, but it cannot be
exactly zero.

\begin{center} \leavevmode \epsfbox{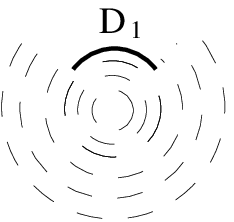} \end{center}

\vskip .1cm
\noindent
{\small {\bf Figure 1.~ Renninger's experiment.} The photon spherical wave is
  modified by  the scintillation detector $D_1$ in spite of the fact that
  it detects nothing.}
\vskip .3cm

Dicke's paradox  is the apparent non-conservation of
 energy in a Renninger-type experiment. He considered an atom in a
 ground state inside a potential well. Part of the well was
 illuminated by a beam of photons. A negative result experiment was
 considered in which no scattered photons were observed, see Fig.~2.
 The atom changed its state from the ground state to some 
 superposition of energy eigenstates (with a larger expectation value of the
 energy) in which the atom does not occupy the part of the well
 illuminated by the photons, while photons (the measuring device)
 apparently have not changed their state at all, and he asked: ``What is
 the source of the additional energy of the atom?!''
 \vskip .3cm

\begin{center} \leavevmode \epsfbox{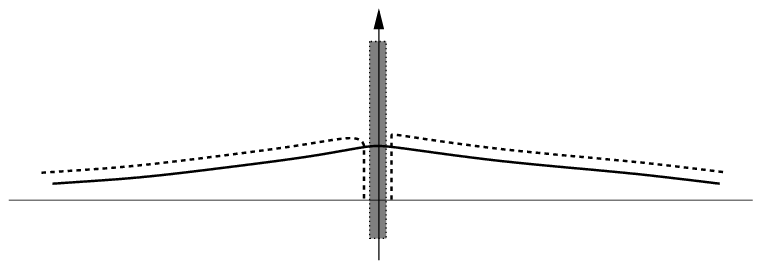} \end{center}

\noindent 
 {\small {\bf Figure 2.~~} {\bf  Dicke's Experiment.} The ground state of a
   particle in the potential well (solid line) is changed to a more energetic
   state (dashed line) due to short radiation pulse, while the quantum state of
   the photons in the pulse remaines unchanged.}
\vskip .3cm 
 Careful analysis \cite{Dick2,Gold} (in part, made by Dicke himself)
 shows that there is no real paradox with conservation of energy,
 although there are many interesting aspects in the process of an
 ideal measurement \cite{Pearl}. One of the key arguments is that the
 photon pulse has to be well localized in time and, therefore, it must
 have a large uncertainty in energy.

 In the Renninger argument, the
 paradox exists only in the formalism of quantum mechanics and,
 moreover, only within some interpretations of quantum theory. The
 ``change'' which occurred ``without interaction'' is the change of a
 quantum state. Adopting the interpretation according to which the
 quantum state does not have its own ``reality'', but is a description
 of our knowledge about the object, removes the paradox completely: of
 course, negative result experiments provide us with some information
 which, consequently, changes our knowledge about the object, i.e., the quantum state of the object.
 
 The IFM of Elitzur and Vaidman which will be discussed in the next
 section is not concerned with changing the object without
 interaction, but with obtaining information about the object without
 interaction. The negative results experiment of Renninger and Dicke
 also provide some information without interaction. We learn, without
 interaction with the object, where the object {\it is not}. This is
 not too surprising: if the object is not in the vicinity of the
 detector, then the detector does not interact with it. We even can
 get information where the object $is$ in this manner provided we have
 {\it prior information} about the state of the object. If it is known
 in advance that the object is somewhere inside two places and it was
 not found in one, obviously, we then know that it is in the second
 place. But this has a trivial classical counterpart: If it is known
 in advance that the object is in one of two separate boxes and we
 open one and do not see it there, then, obviously, we know that it is
 in the second box, and we have not interacted with the object.

\section{THE ELITZUR-VAIDMAN INTERACTION-FREE MEASUREMENTS}
\label{inter}

In the EV IFM paper the following question has been considered:
\begin{quotation}
    Suppose there is an object such that {\em any} interaction with it
  leads to an explosion. Can we locate the object without exploding
  it? 
\end{quotation}

The EV method is based on the Mach-Zehnder interferometer.  A photon
(from a source of single photons) reaches the first beam splitter
which has a transmission coefficient ${1\over2}$.  The transmitted and
reflected parts of the photon wave are then reflected by the mirrors
and finally reunite at another, similar beam splitter, see Fig.~3{\it a}.
Two detectors are positioned to detect the photon after it passes
through the second beam splitter.  The positions of the beam splitters
and the mirrors are arranged in such a way that (because of
destructive interference) the photon is never detected by one of the
detectors, say $D_2$, and is always detected by $D_1$.

This interferometer is placed in such a way that one of the routes of
the photon passes through the place where the object (an ultra-sensitive bomb) might be
present (Fig.~3{\it b}).  A single photon passes through the system.
There are three possible outcomes of this measurement: i)~explosion,~
ii)~detector $D_1$ clicks, iii)~detector $D_2$ clicks.  If 
detector $D_2$ clicks (the probability for that is ${1\over4}$), the
goal is achieved: we know that the object is inside the interferometer
and it did not explode.

\begin{center} \leavevmode \epsfbox{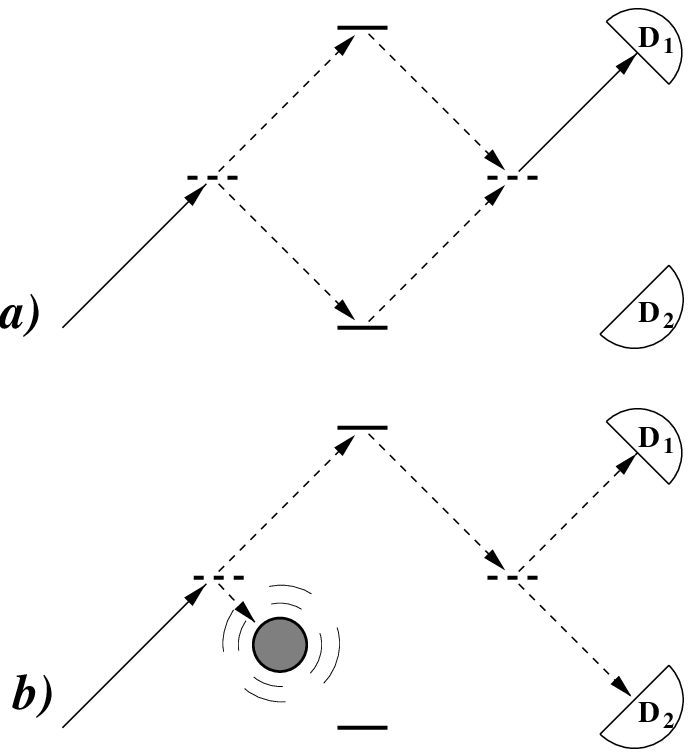} \end{center}
\noindent 
{\small {\bf Figure 3.~~}(a) When  the interferometer is
  properly tuned, all photons are detected by $D_1$ and none reaches
  $D_2$. \hfill\break (b) If the bomb is present, detector $D_2$ has the
  probability 25\% to detect the photon  sent through the
  interferometer, and in this case we know that the bomb is inside the
  interferometer without exploding it.  }

\vskip .4cm

The EV method solves the problem which was stated above. It allows
finding with certainty an infinitely
sensitive bomb without exploding it. The bomb might explode in the
process, but there is at least a probability of 25\% to find the bomb
without the explosion. ``Certainty'' means that when the process is successful ($D_2$
clicks), we know for sure that there is something
inside the interferometer. (A modification of the EV IFM which employs the quantum Zeno effect allows to reduce the probability of the explosion to an arbitrarily small value \cite{Kwi95}.) 

In an earlier paper, Renninger \cite{Ren53} considered an experimental setup which is
almost identical to that of the EV IFM: a Mach-Zehnder
interferometer tuned to have a dark output towards one of the
detectors. However, he never regarded his experiment as a
measurement on an object which was inside the interferometer:
Renninger's argument, as in the experiment described in Fig. 1, was
about ``interaction-free'' changing the state of the photon. Renninger
has not asked the key question of the EV IFM: How to get information
in an interaction-free manner.

The basic paradox of the EV IFM can be presented in the following way.
The only interaction of the bomb with an external world is through its
explosion. Nevertheless, the EV scheme allows finding the object
without the explosion. It is different from the trivial way of the
Renninger-Dicke IFM when we know before the  experiment that the bomb is
somewhere in a particular region and then, not finding it on part of
the region, tells us that it is in the remaining part. The EV method
works even if we do not have  {\it any} information about the location or
even the existence of the object prior to the measurement.

The weakness of this paradox can be seen in sentences: ``Suppose there is an object such that {\em any} interaction with it
  leads to an explosion,''  ``The only interaction
of the bomb with an external world is through its explosion.''
Quantum mechanics precludes existence of such objects. Indeed, a good
model for an ``explosion'' is an inelastic scattering \cite{Gesz}. The
Optical Theorem \cite{Land} tells us that there cannot be an inelastic
scattering without some elastic scattering. Thus, if it were such an
object, the quantum experiment of the EV IFM would find it without
interaction, but since quantum theory ensures that there are no
objects like this, it also avoids the paradox. Not exactly. The
EV method is still very paradoxical: it employs the explosion for
detection but it does not cause the explosion (at least in some
cases.) The task of the IFM can be rephrased in the following way:
\begin{quotation}
    Suppose there is an object such that a  particular interaction with it
  leads to an explosion. Can we locate the object without exploding
  it using this interaction? 
\end{quotation}

\section{SECOND PARADOX OF THE EV IFM:  MEASUREMENT ``WITHOUT TOUCHING''}
\label{int}

Suppose there is a place in the Universe that no particle, no light,
nothing whatsoever visited, i.e. no particle passed through
this place, no particle went to this place and was stopped there.
Suppose also that nothing came out of this place: no particle, no
field, no source of potential observable through the Aharonov-Bohm type
effect, nothing whatsoever. It seems that in this case we cannot know:
Is there  something in this place? If, however, we put the mirrors 
of the EV interferometer around this place such that one of the arms
of the interferometer crosses it and send through the interferometer
a single photon which ends up in $D_2$, then we know that there is
something there. Moreover, if we later find out that this
``something'' is a nontransparent object then we can claim that we have found
it without ``touching'': nothing was in the vicinity of the object. 

 This claim,
again, has to be taken in an appropriate context. It has the same
justification as  Wheeler's delayed choice experiment analysis \cite{Whee}.
 One of the ``choices'' of Wheeler's delayed-choice experiment is an
 experiment with a Mach-Zehnder interferometer in which the second beam
 splitter is missing, see Fig.~4. In a run of the experiment with a
 single photon detected by $D_2$, it is usually accepted that the
 photon had a well defined trajectory: the upper arm of the
 interferometer. In contrast, according to the von Neumann approach,
 the photon was in a superposition inside the interferometer until the
 time when one part of the superposition reached  detector $D_2$ (or
 until the time the other part reached  detector $D_1$ if that
 event was earlier). At that moment the wave function of the photon
 collapses to the vicinity of $D_2$. 
 
 The justification of Wheeler's claim that the photon detected by
 $D_2$ never was in the lower arm of the interferometer is that,
 according to the quantum mechanical laws, we cannot see any physical
 trace from the photon in the lower arm of the interferometer. This is
 true if (as it happened to be in this experiment) the photon from the
 lower arm of the interferometer cannot reach the detector $D_2$. The fact
 that there cannot be a physical trace of the photon in the lower arm of
 the interferometer can be explained in the framework of the two-state
 vector formulation of quantum mechanics \cite{ABL,AV90}.  This
 formalism is particularly suitable for this case because we have a pre-
 and post-selected situation: the photon was post-selected at $D_2$.
 Thus, while the wave function of the photon evolving forward in time
 does not vanish in the lower arm of the interferometer, the
 backward-evolving wave function does.  Vanishing of one of the waves
 (forward or backward) is enough to ensure that the photon cannot
 cause any change in local variables of the lower arm of the
 interferometer.

\begin{center} \leavevmode \epsfbox{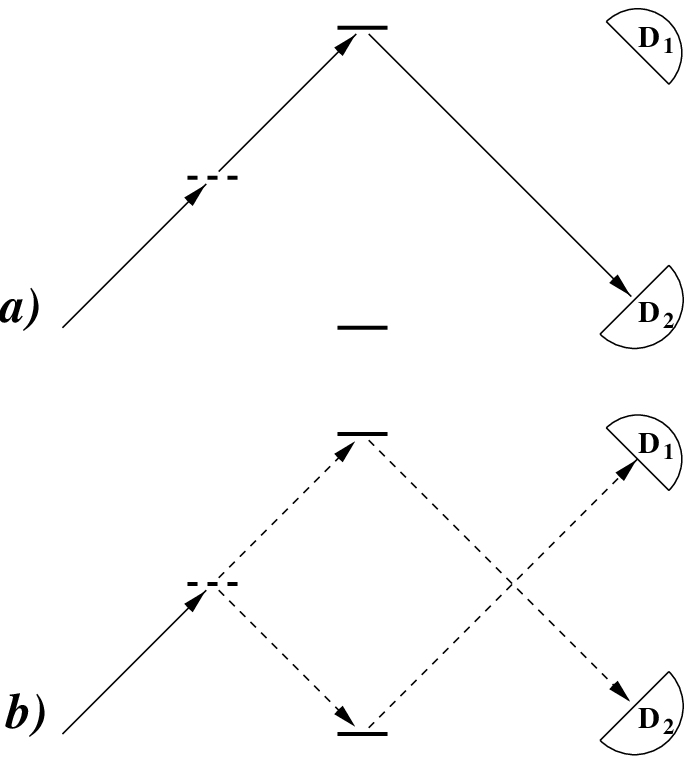} \end{center}

\noindent 
 {\small {\bf Figure 4.~~} (a) The ``trajectory'' of the photon in the
   Wheeler experiment given that $D_2$ detected the photon as it
   is usually described. The photon cannot leave any physical trace
   outside its ``trajectory''.  \hfill\break (b) The ``trajectory''
   of the quantum wave of the photon in the Wheeler experiment
   according to the von Neumann approach. The photon remains in a
   superposition until the collapse which takes place when one of the
   wave packets reaches a detector.  }

 \vskip .4cm

 In the EV IFM we have the same situation. If there is an object in
 the lower arm of the interferometer, see Fig.~3{\it b}, the photon
 cannot go through this arm to detector $D_1$. This is correct if the
 object is such that it explodes whenever the photon reaches its
 location and we have not observed the explosion. Moreover, this is
 also correct in the case in which the object is completely
 non-transparent and it blocks the photon in the lower arm eliminating
 any possibility of reaching $D_1$.  Even in this case, when the
 object does not explode on touching, we can claim that we locate the
 object ``without touching''. This claim is identical to the argument
 according to which the photon in Wheeler's experiment went solely
 through the upper arm.

 In the framework of the two-state vector approach this is explained
 in the following way.  The forward-evolving quantum state vanishes in
 the lower arm of the interferometer beyond the location of the
 object, while the backward-evolving wave function vanishes before the
 location of the object. Thus, at every point of the lower arm of the
 interferometer one of the quantum states vanishes. This ensures that
 the photon cannot make any physical trace there. Note, that the
 two-state vector formalism itself does not suggest that the photon is
 not present at the lower arm of the interferometer; it only helps to
 establish that the photon does not leave a trace there. The latter is
 the basis for the claim that in some sense the photon was not there.

\section{THE PENROSE  INTERACTION-FREE MEASUREMENTS}
\label{intPen}

The task of the EV IFM is to find the location of an object without
interaction. Penrose proposed to use a similar idea for testing some property of an object
without interaction \cite{Pen}. The object is again a bomb which explodes when
anything, even a single photon, ``touches'' its trigger device. Some of the
bombs are broken (they are duds) in a particular way: their trigger
device is locked to a body of the bomb and no explosion and no motion
of the trigger device would happen when it is ``touched''. 
Again, the paradox is that any touching of a good bomb leads to an
explosion. How can we test the bomb without exploding it?

In the Penrose version of IFM, the bomb plays the role of one mirror of
the interferometer, see Fig.~5. It has to be placed in the correct
position. We are allowed to do so by holding the body of the bomb. However, the
uncertainty principle puts limits on placing the bomb in its place
before the experiment \cite{Pen-Vai}. Only if the position of the bomb
(in fact, what matters is the position of a dud) is known exactly, the
limitations are not present.  In contrast, in the EV IFM the bomb need
not be localized prior to the measurement: the IFM localizes it by
itself.

\begin{center} \leavevmode \epsfbox{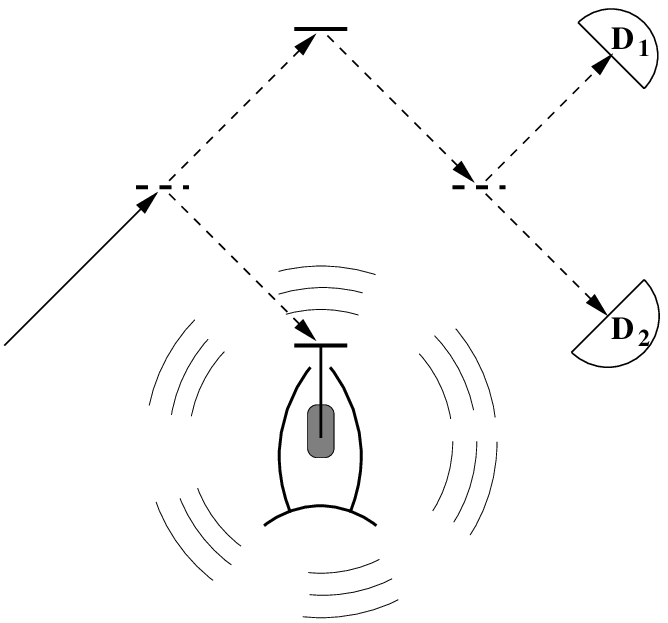} \end{center}
\noindent 
{\small {\bf Figure 5.} {\bf The Penrose bomb-testing device.} The mirror of the good bomb
  cannot reflect the photon, since the incoming photon causes an
  explosion. Therefore, $D_2$ sometimes clicks. (The mirror of a dud
  is connected to the massive body, and therefore the interferometer
  ``works'', i.e. $D_2$ never clicks when the mirror is a dud.)  }

\section{ Interaction-free localization of a quantum object}
\label{IFML}

When the EV IFM is applied to a quantum object spread out in space, it
collapses the spatial wave function  without 
changing the state of internal variables \cite{EV93}. Let us discuss two aspects of such experiments.

First, in order to see the difference between the Renninger-Dicke IFM and the EV IFM
 more vividly, let us consider an application of the EV method to
 Dicke's experimental setup.  Instead of the light pulse we send a
 ``half photon'': We arrange the EV device such that  one arm of the
 Mach-Zehnder interferometer passes through the location of the
 particle, see Fig.~6.  Then, if detector $D_2$
 clicks, the particle is localized in the interaction region.
 
 In both cases (the Renninger-Dicke IFM and this EV IFM) there is a
 change in the quantum state of the particle without apparent scattering
 of the photon by the particle.  However, the situations are quite
 different. In the original Dicke's experiment we can claim that the
 dashed line of Fig.~2.  is the state of the particle after the
 experiment only if we have prior information about the state of the
 particle before the experiment (solid line of Fig.~2.) In contrast,
 in the EV modification of the experiment, we can claim that a
 particle is localized in the vicinity of the interaction area (dashed
 line of Fig.~6.) even if we had no prior information about the
 state of the particle.

\begin{center} \leavevmode \epsfbox{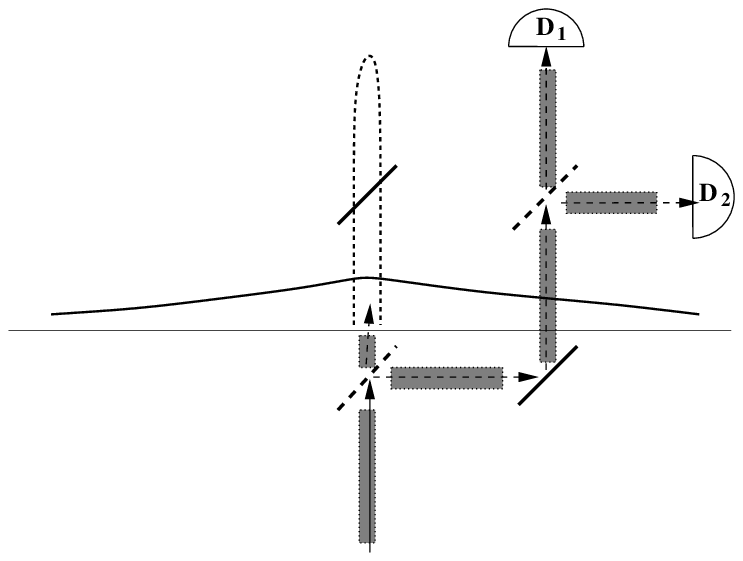} \end{center}

\noindent 
{\small {\bf Figure 6.} {\bf The EV modification of Dicke's
  Experiment.} The ground state of a particle in the potential well (solid
  line) is changed to well a localized state (dashed line) when the
  photon is detected by the detector $D_2$.}

\vskip .3cm

 The second aspect of the EV IFM applied to quantum objects is that 
the argument according to which   the measurement was
performed without a  photon being in the vicinity of the object, encounters
a  subtle difficulty: it might be the case that we perform the
procedure of the IFM, obtain the photon click at $D_2$, but,
nevertheless, the photon was {\it  with certainty} in the area of interaction.

First, let us repeat the argument which led us to think that the
photon was not there.
 If $D_2$ clicks,
we can argue that the particle had to be on the way of the photon in
the left arm of the interferometer (in the right arm the trajectories
do not intersect), otherwise, it seems that we cannot explain the
arrival of the photon to the ``dark'' detector $D_2$. If the particle
was on the way of the photon in the left arm of the interferometer we
can argue that the photon was not there, otherwise we had to see the
explosion. Therefore, the photon went through the right arm of the
interferometer and it was not present in the left arm of the interferometer. 

The persuasive argument of the previous paragraph is wrong! Not just
the semantic point discussed above, i.e., that the quantum wave of the
photon in the left arm of the interferometer in the part before the
``meeting point'' with the particle was not zero, is incorrect.  It is
wrong to say that the photon was not in the left arm even in the part
{\it beyond} the meeting point with the atom. In an unambiguous
operational sense it is wrong to say that in the experiment in which
$D_2$ clicks, the probability to find (in a non-demolition way) the
photon  in the left arm of the interferometer after the meeting point
with the atom is zero. The photon {\it can} be found in the left arm of the interferometer! A particular way to achieve this  is  discussed 
in the next section.

\section{Hardy's paradox}
\label{hard}

Hardy considered ``nested
interaction-free measurements'' \cite{Hardy}.  The particle is in a
superposition of two wave packet inside its own Mach-Zehnder
interferometer (see Fig.~7.)  If $D_2$ (for the photon) clicks, the
particle is localized inside the interferometer. If we assume that
before the experiment the whole volume of the interferometer except
the ``meeting place'' $W$ which we want to test was found empty, we can
claim that the click of $D_2$ localizes the particle inside $W$.
However, the particle plays the role of the photon of another IFM (we
can consider a gedanken situation in which the particle which explodes
when the photon reaches its location can, nevertheless, be manipulated
by other means). If this other IFM is successful (i.e.  ``$D_2$'' for
the particle clicks) then the other observer can claim that she
localized the  photon of the first experiment in the ``meeting
place'' $W$, i.e. that the photon passed through the lower arm of the
interferometer on its way to $D_2$.

\begin{center} \leavevmode \epsfbox{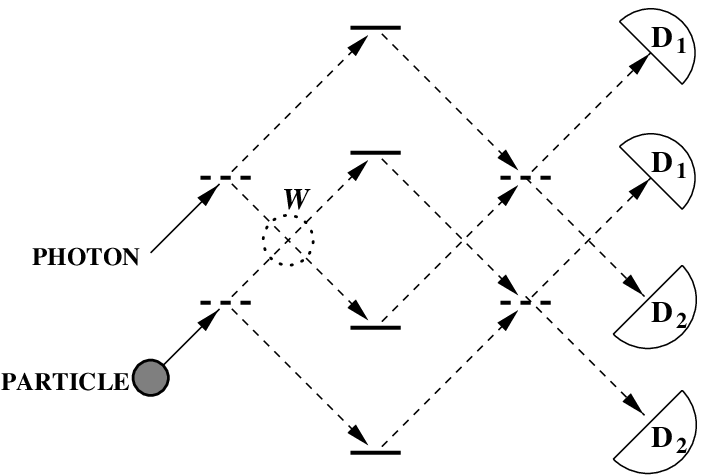} \end{center}
\noindent 
{\small {\bf Figure 7.~~} {\bf Hardy's Paradox.} Two interferometers
  are tuned in such a way that, if they operate separately,  there
  is a complete destructive interference towards detectors $D_2$. The
  lower arm of the photon interferometer intersects the upper arm of
  the particle interferometer in $W$ such that the particle and the
  photon cannot cross each other.  When the photon and the particle
  are sent together (they reach $W$ at the same time) then there is a
  nonzero probability for clicks of both detectors $D_2$. In this case
  one can infer that the particle was localized at $W$ and also that
  the photon was localized at $W$.  However, the photon and the
  particle were not present in $W$ together. This apparently
  paradoxical situation does not lead to a real contradiction because
  all these claims are valid only if tested separately.}

\vskip .3cm

 Paradoxically, both claims are true (in the operational
sense): the first experiment localizes the particle in  $W$, and the
second, at the same time, localizes the single photon there.  Both
claims are true separately, but not together: if we would try to find
both the photon and the particle in $W$, we will fail with certainty.
Such peculiarities take place because we consider a pre- and
post-selected situation (the post-selection is that in both
experiments detectors $D_2$ click) \cite{Har-Vai}.

In spite of this peculiar feature, the experiment is still
interaction-free in the following sense.  If somebody would test the
success of our experiment for localization of the particle, i.e. would
measure  the location of the particle shortly after the ``meeting time'' between
the particle and the photon, then we know for sure that she would
find the particle in the left arm of the interferometer and,
therefore, the photon cannot be there. Discussing the issue of the
presence of the particle with her, we can correctly claim that in our
experiment the photon was not in the vicinity of the particle.  Again,
at the end of the EV IFM procedure for a quantum object we cannot
claim that the photon we used was not in the vicinity of $W$.  But we
can still claim that we have localized the particle there without the
photon being in $W$. To localize means that if tested it must be found
there, and if tested, the photon was not there.  However, if, instead
of measuring the position of the particle after the meeting time, she
finds the particle in a particular superposition, she can claim with
certainty that the photon was in $W$. (Compare this with {\it
  deterministic quantum interference experiments} \cite{APP}).

\section{Resolution of the paradoxes in the framework of the Many-Worlds Interpretation}
\label{MWI}

I want to argue that the paradoxes presented above are
resolved, or at least appear less paradoxical in the framework of the
Many-Worlds Interpretation (MWI) \cite{MWI}. The MWI itself has
several
interpretations, some of which are conceptually different. I 
 take the view
in which we have {\it one} physical universe which incorporates many
(subjective) worlds \cite{Va-mwi}. The physical universe is described by one wave
function evolving deterministically  according to the Schr\"odinger
equation. This wave function can be decomposed into a superposition of
many states, each corresponding to a different story. One of the stories is the
world as {\it you}, the reader of this paper, know it. What we
perceive is just a small part of what {\it is} in the universe.
The laws of physics relate to the whole universe, and it is not
surprising that considering   only a part of it
leads to paradoxical situations. Considering the physical universe, i.e. all worlds together,
 resolves the paradoxes \cite{Va-par}.

In the framework of the MWI it is not true that we get
information about the region without anything being there. It is not
true that we find the bomb without an explosion.
The photon which we sent into the interferometer was there, but --  in
another world. In our experiment three  worlds (three different
stories) appear:\hfil\break
(i)~there is an explosion,\hfil\break
(ii)~detector $D_1$ clicks,\hfil\break
(iii)~detector $D_2$ clicks.\hfil\break
  Obtaining information in the world (iii)
without any object being in the region became possible because in the
world (i) a photon was in the region and it caused the explosion.

The EV IFM allows to find an object in a particular place without
visiting the place. However, it does not allow to find out that the
place is empty without being there. The MWI explains why it is so.
 If the place is empty, then in the EV procedure there is only one
world, the one we are aware of, so obtaining information about
the region without being there is on the level of the whole universe.
Our physical intuition correctly tells us
that such a situation is impossible.

What I can see in common between the Renninger-Dicke IFM and the EV
IFM is that in the framework of the many-worlds interpretation in both
cases we can see the ``interaction'': radiation of the scintillator in the
Renninger experiment or explosion of the bomb in the EV experiment,
but these interactions take place in the ``other'' branch, not in the
branch we end up discussing the experiment. (In an attempt to avoid
adopting the many-worlds interpretation such interactions are considered as
{\it counterfactual}; see \cite{Pen} p.240 and \cite{Mi-Jo}.

To conclude this paper, let me express my attitude to quantum
paradoxes. Paradoxes in physics are very important: they lead to new theories.
There are numerous paradoxes in quantum mechanics, but, in my view,
none of them is a real paradox which will lead to new physical
laws. The quantum mechanical paradoxes do not follow from
incorrectness or incompleteness of quantum theory, but from
inappropriate classical intuition which people developed during
thousands of years when quantum phenomena were not observed and thus
no one had reason to believe in quantum mechanics. The role of quantum 
paradoxes is not to lead to new theories but to lead to the development of 
new intuition about our world. Here, I probably will be in a minority: 
I prefer to believe that there is no conceptually new physics which we do
not know yet, I prefer the feeling that we, basically, do  understand our world.

\vspace{.3cm}
 \centerline{\bf  ACKNOWLEDGMENTS}
 
 It is a pleasure to thank Yakir Aharonov, Berge Englert, Harry Paul,
 and Philip Pearle for helpful discussions and Rodolfo Bonifacio for
 promoting quantum paradoxes to be a contemporary scientific topic.
 This research was supported in part by grant 471/98 of the Basic
 Research Foundation (administered by the Israel Academy of Sciences
 and Humanities) and the EPS grant of Research Council GR/N33058.

\end{multicols}

\end{document}